\documentclass[fleqn,12pt]{article}

\usepackage[dvips]{epsfig}

\begin{document}

\title{Complex functions as lumps of energy \\
\emph{Funciones complejas como c\'umulos de energ\'{\i}a}}

\author{$\Re$J Cova\footnote{rcova@luz.ve} \\ Dept
de F\'{\i}sica, FEC \\ La Universidad del Zulia \\ Apartado 15332 \\
Maracaibo 4005-A \\ Venezuela \\
\and C. Uberoi\footnote{cuberoi@math.iisc.ernet.in} \\
Dept of Mathematics and\\
Supercomputer Education and\\ Research Center \\
Indian Institute of Science \\
Bangalore 560 012 \\ India}              

\maketitle


\abstract{We present an application of the basic mathematical
concept of complex functions as topological solitons,  
a most interesting area of research in physics. 
Such application of complex theory is virtually
unknown outside the community of soliton researchers.

\vspace{.2in}

\emph{Presentamos una aplicaci\'on del concepto matem\'atico de
funciones complejas como solitones topol\'ogicos, una interesante
\'area de investigaci\'on en f\'{\i}sica. Dicha aplicaci\'on de la teor\'ia
compleja es pr\'acticamente desconocida fuera del c\'{\i}rculo de
investigaci\'on solit\'onica.}}          

\section*{Introduction \label{sec:intro}}

The complex variable $z=x+iy$, where $x,y \in \Re$ and $i=\sqrt{-1}$, is one
of the most familiar and useful concepts in mathematics, with a very large number
of well-documented applications in science.

Over the past few years some interesting nonlinear models in physics have
received a lot of attention, models bearing the so-called solitons or 
energy `lumps'. Some of these models exemplify
yet another important application of                                
complex functions, with functions
as simple as $f(z)=z$ describing a soliton configuration.
Unfortunately, despite the vast literature dealing with complex 
analysis plus applications, one finds no mention of the  
starring role of $z$ as a soliton. Reference to
such an extraordinary role is found only in highly specialised research 
books and journals, hence the existence of $z$ as a soliton field 
is practically unknown outside the group of specialists in the
area.

Using the nonlinear sigma $O(3)$ (or $CP^1$) model in two spatial dimensions, the
present work illustrates the context in which $z$ stands for a lump of energy. This 
fact is most remarkable and, given the growing importance of solitons in physics, we
believe that more physicists should know about it. They will find
this fresh utility of complex variables quite appealing.        

\section{Complex theory}

Complex theory is a very important branch of mathematics.  As a brush-up
we just recall that many integrals given in real form are easily
evaluated by relating them to complex integrals and using the powerful
method of contour integration based on Cauchy's theorem. In fact, the
basis of transform calculus is the integration of functions of a complex
variable. And intersections between lines and circles, parallel or 
orthogonal lines, tangents, and the like usually become quite simple 
when expressed in complex form.

Familiarity with the complex numbers starts early, when at high school the basics
of $z$ are taught. Then in college algebra/calculus one learns some more about
complex variables, with immediate applications to problems in both physics and
engineering like electric circuits and mechanical vibrating systems. Later on
complex holomorphic (analytic) functions are introduced, and then applied to a
variety of problems:  heat flow, fluid dynamics, electrostatics and magnetostatics,
to name but few. 

The concept of analyticity is extremely important. 
Many physical quantities are represented by functions $f(x,y), \; g(x,y)$ 
connected
by the relations 
\(
\partial_x f=\partial_y g, \quad 
\partial_y f=-\partial_x g, 
\)
where \( \partial_x f=\partial f/\partial x, \; \) \emph{etc}. 
It turns out that $f$ and $g$ may be considered
as the real and imaginary parts of a holomorphic function
$h$ of the complex variable $z$:
\begin{equation}
h(z)=f+ig.
\label{eq:h}
\end{equation}
The equations linking $f$ and $g$ are the Cauchy-Riemann conditions
for $h(z)$ being holomorphic, and can be written compactly as
\begin{equation}
\partial_x h=-i \partial_y h.
\label{eq:cauchy}
\end{equation}      
When $h$ is a function of $\bar{z}=x-iy$, the complex conjugate of $z$,
the condition (\ref{eq:cauchy}) reads $\partial_x h=i \partial_y h$,
and $h(\bar{z})$ is said to be anti-holomorphic \cite{ahlf}.   

We hereby show how functions of the type (\ref{eq:h}) describe solitons,
giving yet another fundamental, if little known, application of analytic
complex functions.

\section{Solitons}

Nonlinear science has developed strongly over the past 40 years,
touching upon every discipline in both the natural and social
sciences. Nonlinear systems appear in mathematics, physics, chemistry, 
biology, astronomy, metereology, engineering, economics and many more
\cite{lam,lak}.

Within the nonlinear phenomena we find the concept of `soliton'.  It has got
some working definitions, all amounting to the following picture: a travelling wave
of semi-permanent lump-like form. A soliton is a non-singular solution of a
non-linear field equation whose energy density has the form of a lump localised in
space. Although solitons arise from nonlinear wave-like equations, they
have properties that resemble those of a particle, hence the suffix \emph{on} to
covey a corpuscular picture to the \emph{soli}tary wave.
 Solitons exist as density waves in spiral galaxies, as lumps in the
ocean, in plasmas, molecular systems, protein dynamics, laser pulses
propagating in solids, liquid crystals, elementary particles, nuclear
physics...

According to whether the solitonic field equations can be solved or not, solitons
are said to be integrable or nonintegrable. Given the limitations to analitycally
handle nonlinear equations, it is not surprising that integrable solitons are
generally found only in one dimension. The dynamics of integrable solitons is
quite restricted; they usually move undistorted in shape and, in the event of a
collision, they scatter off undergoing merely a phase shift. 

In higher dimensions the dynamics of solitons is far richer, but now we
are in the realm of nonintegrable models. In this case analytical
solutions are practically restricted to static configurations and Lorentz
transfomations thereof. (The time evolution being 
studied via numerical simulations and other approximation techniques.) A
trait of nonintegrable solitons is that they carry a conserved
quantity of topological nature, the topological charge --hence the
designation \emph{topological solitons}. Entities of this kind 
exhibit interesting stability and scattering processes, including 
soliton annihilation which can occur when lumps with opposite topological
charges (one positive, one negative) collide. For areas like 
nuclear/particle physics such dynamics is of great relevance.

Using the simplest model available, below we illustrate the emergence of topological 
solitons and their representation as complex functions.

\section{The planar $O(3)$ model}

Models in two dimensions have a wide range of applications. In physics
they are used in topics 
that include Heisenberg ferromagnets, the quantum Hall effect,
superconductivity, nematic crystals, topological fluids, vortices and
solitons. 
Some of these models also appear as low dimensional analogues of 
forefront non-abelian gauge field 
particle theories in higher dimensions, 
an example being the Skyrme model of nuclear physics 
\cite{sky1,sky2}.

One of the simplest such systems is the $O(3)$ or $CP^1$ sigma 
model in (2+1) dimensions (2 space, 1 time). It involves 
three real scalar fields $\phi_j$ ($j$=1,2,3) functions of the 
space-time coordinates $(t,x,y)$ \cite{raja,wjz}. The model is defined by the
Lagrangian 
density
%
\begin{equation}   
{\cal L}=\frac{1}{4} \sum_{j=1}^{3} 
[ (\partial_{t}\phi_j)^2 - (\partial_{x}\phi_j)^2
- (\partial_{y}\phi_j)^2 ]
  \label{eq:lagphi}
\end{equation}
where the fields, compactly written as the vector in field space 
$\vec{\phi} \equiv (\phi_1,\phi_2,\phi_3)$, are constraint to lie on 
the unit sphere:
%
\begin{equation}
S^{(\phi)}_2=\{\vec{\phi}: \vec{\phi}^2=1\}.
\label{eq:constraint}
\end{equation}           
The Euler-Lagrange field equation derived from
(\ref{eq:lagphi})-(\ref{eq:constraint}) has no known analytical solutions 
except for the static case, which equation reads
%
%
\begin{equation}
\nabla^2 \vec{\phi} -
(\vec{\phi}.\nabla^2 \vec{\phi})\vec{\phi} = \vec{0} 
\quad [ \nabla^2 \equiv \partial_x^2 + \partial_y^2].
\label{eq:staticfieldeqo3}
\end{equation}    

The $CP^1$ solitons are non-singular solutions of (\ref{eq:staticfieldeqo3}).
Without the constraint (\ref{eq:constraint}) the said equation
would reduce to 
$
\nabla^2 \vec{\phi} = \vec{0},
$
whose only non-singular solutions are constants. The condition 
(\ref{eq:constraint}) leads to the second term in  
(\ref{eq:staticfieldeqo3}), equation which
does yield non-trivial non-singular solutions as we will
later see. 

Solitons must also be finite-energy configurations. From  
(\ref{eq:lagphi}) we readily get the static energy 
%
\begin{eqnarray}
E &=& \int \frac{1}{4} \sum_{j=1}^{3} 
      [(\partial_{x}\phi_j)^2+(\partial_{y}\phi_j)^2]\, d^2 x 
                                \nonumber \\
  &=& \int \frac{1}{4} \sum_{j=1}^{3} 
       (\nabla \phi_j)(\nabla \phi_j)\, d^2 x 
\quad [\nabla \equiv (\partial_x,\partial_y)] \nonumber \\
  &=& \frac{1}{4} \int 
      (\nabla \vec{\phi}).(\nabla \vec{\phi})\, rdrd\theta
\quad (\mbox{in polar coordinates} \; r, \theta).
\label{eq:e}
\end{eqnarray}
We ensure the finiteness of $E$ by taking the boundary condition
%
\begin{equation}        
\lim_{r \rightarrow \infty} \vec{\phi}(r,\theta)
\rightarrow \vec{\phi}_{0} \quad (
\mbox{a constant unit vector independent of} \; \; \theta), 
\label{eq:bc}
\end{equation}       
since 
the integrand in (\ref{eq:e}) will thus tend to zero at spatial 
infinity :
%
\begin{equation}
\lim_{r \rightarrow \infty} r |\nabla \vec{\phi}|=
\lim_{r \rightarrow \infty} r 
\sqrt{(\partial_r \vec{\phi})^2
+ (\frac{1}{r} \partial_{\theta} \vec{\phi})^2}
\rightarrow  0.
\label{eq:bc2}
\end{equation}
%
\subsection{The complex plane}
%
We are thus considering the
model in the $x-y$ plane with a point at infinity, 
\emph{ie}, the  \underline{extended 
complex plane} which is topologically equivalent to the two-sphere
$S^{(x)}_2$.
The finite energy
configurations
are therefore fields $\vec{\phi}$ defined on 
$\Re_2 \cup \{ \infty \} \cong S^{(x)}_2$ 
and taking values on $S^{(\phi)}_2$. In other words, our 
finite-energy fields are harmonic maps 
of the form $S^{(x)}_2 \rightarrow S^{(\phi)}_2$ \cite{eells}.

We may imagine the coordinate space $S^{(x)}_2$ as made of \emph{rubber} and
the field space $S^{(\phi)}_2$ as made of \emph{marble}; the map $\vec{\phi}$
constrains the rubber to lie on the marble. Then with each point
$(x,y)$ in the rubber we have a quantity 
$$
\vec{\tau}=\nabla^2 \vec{\phi} -
(\vec{\phi}.\nabla^2 \vec{\phi})\vec{\phi}
$$
at the point $\vec{\phi}$ in marble representing the tension 
in the rubber at that point. Thus the map is harmonic if 
and only if $\vec{\phi}$ constrains the rubber to lie on the 
marble in a position of elastic equilibrium, $\vec{\tau}=\vec{0}, \;$ 
which is just  equation (\ref{eq:staticfieldeqo3}). These are our 
finite-energy configurations, of which the soliton solutions are a subset.

\subsection{Topological charge}

In general, 
as the coordinate $z=(x,y)$ ranges over the sphere $S^{(x)}_2$ once,
the coordinate $\vec{\phi}=(\phi_1,\phi_2,\phi_3)$ ranges over 
$S^{(\phi)}_2$ $N$ times. This winding number is called the 
topological charge in soliton parlance, and classifies the maps 
$S^{(x)}_2 \rightarrow S^{(\phi)}_2$ into sectors
(homotopy classes); maps within one sector are equivalent in that they
can be obtained from each other by continuos transformations. 

An expression for the topological charge is obtained by expanding 
the coordinates $\phi_j$ of the area element 
of $S^{(\phi)}_2$ in terms of coordinates $(x,y)$ in $S^{(x)}_2$, and
integrating off. In plainer language, from the college formula that computes
the flux of a vector $\vec{A}$ through a region $D$ of a surface $S$ :
\begin{equation}
\int_D \vec{A}.\widehat{\mathbf{n}} dS \quad 
[\widehat{\mathbf{n}} \; \mbox{a normal unit vector}],
\label{eq:flux}
\end{equation}
the topological charge $N$ follows by calculating the flux of 
$\vec{A}=\frac{\textstyle{N}}{\textstyle{4 \pi a^2}} \vec{\phi}$ through the sphere
$S_2^{(x)}$
of radius $a=1$:
\begin{displaymath}
\int_D \vec{A}.\widehat{\mathbf{n}} dS \rightarrow 
\int_{S_2^{(x)}} \frac{N}{4 \pi a^2} \vec{\phi}.\vec{\phi} \, dS = N.
\end{displaymath}

Notably, a field with topological charge $N$ describes precisely a system 
of $N$ solitons.  

In order to actually find charge-$N$ finite-energy \emph{solutions},
it is convenient to express the
model in terms of one independent complex field, $W$, related to  
$\vec{\phi}$ via the stereographic projection
%
\begin{equation}
W=\frac{\phi_{1}+i\phi_{2}}{1-\phi_{3}}.
\label{eq:relationwphi}
\end{equation}
In this formulation, the topological charge is given by
\begin{equation}
N=\frac{1}{\pi} \int_{S_{2}^{(x)}}
\frac{|\partial_{z}W|^{2}-
|\partial_{\bar{z}}W|^{2}}{(1+|W|^2)^{2}} \, d^2 x,
\quad N \in {\cal Z},
\label{eq:chargew}
\end{equation}         
connected with the energy (\ref{eq:e}) through
\begin{equation}
E \geq 2 \pi |N|.
\label{eq:bound}
\end{equation}

\section{Lumps}

The solitonic solutions we seek correspond to the 
equality in (\ref{eq:bound}) \cite{tak,belavin,woo}. That is, in a given
topological sector $N$ the static solitons of the planar $CP^1$ model
are the configurations whose energy $E$ is an absolute minimum. Combining 
(\ref{eq:chargew}) with $E=2\pi|N|$ we find that solutions
carrying positive or negative topological charge satisfy,
respectively,
\begin{equation}
\partial_{\bar{z}} W=0 \rightarrow \partial_x W = -i \partial_y W,
\label{eq:cauchyz}
\end{equation}
\begin{equation}
\partial_z W=0 \rightarrow \partial_x W = i \partial_y W.
\label{eq:cauchyzbar}
\end{equation}           
But recalling equation (\ref{eq:cauchy}) we
immediately recognise the above equations as
the Cauchy-Riemann conditions for $W$ being a holomorphic function
of $z$ or $\bar{z}$. This is most remarkable.

\begin{figure} 
\mbox{\epsfig{file=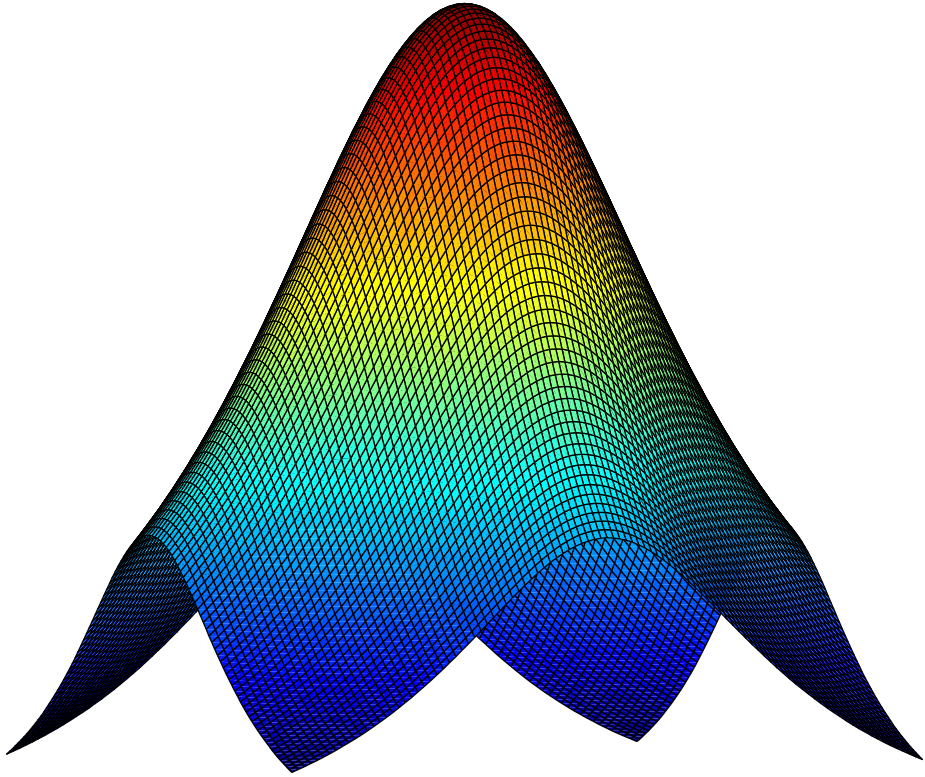}} 
\caption{The energy distribution corresponding to the soliton $W=z$.}
\label{fig:csfig}
\end{figure}

For instance, a single-soliton solution ($N=1$) may be described by  
\begin{equation}
W=z \qquad
[\mbox{note that this satisfies equation (\ref{eq:cauchyz})}];
\label{eq:wz}
\end{equation} 
its energy density distribution is given by
\begin{equation}
{\cal E}=\frac{2}{1+|z|^2}.
\label{eq:ez}
\end{equation}
Plots of (\ref{eq:ez}) reveal a lump of energy localised in space, 
as shown in figure \ref{fig:csfig}. The same energy corresponds to 
$W=\bar{z}$, which has $N<0$ and sometimes is referred to as an
anti-soliton.
                                                          
A more general $N=1$ solution is given by a rational function
\linebreak
$W=\lambda (z-a)/(z-b)$, which we should note is non-singular: 
$W(z=b)=\infty$ corresponds to $\phi_3=1$, the north pole of 
$S_2^{(\phi)}$ according to (\ref{eq:relationwphi}). A prototype solution
for arbitrary $N>0$ is $\lambda (z-a)^N$. 

The dynamics of these structures is studied by numerically
evolving the full time-dependent equation derived from 
(\ref{eq:lagphi}), with the fields $W(z)$ as initial conditions
\cite{leese,cova}.

Sigma $CP^1$-type models have several applications, noteworthy among
them being the Skyrme model in (3+1) dimensions where the
topological solitons stand for ground states of light nuclei, with the
topological charge representing the baryon number.               

The role of complex functions as topological solitons
deserves widespread attention and should 
not be missing from the modern literature dealing with complex theory
and its applications. 

\newpage
\Large{\bf Acknowledgements} \\

\normalsize
$\Re$J Cova thanks the Third World Academy of Science (TWAS)
for its financial support and the Indian Institute of Science (IISc) for 
its hospitality. The support of \emph{La
Universidad del Zulia} is greatly acknowledged. The authors thank 
Mr A. Upadhyay for helpful conversations.
                                    
\newpage

\end{document}